\documentclass[preprint,12pt]{elsarticle}

 \usepackage{graphicx}

\usepackage{amssymb}

\usepackage{rotating}
\usepackage{url}
\usepackage{hyperref}

\journal{Computer Physics Communications}

\begin{document}

\begin{frontmatter}



\title{BAT - The Bayesian Analysis Toolkit}


\author[MPI]{Allen Caldwell\corref{cor2}}
\author[MPI]{Daniel Koll\'ar}
\address[MPI]{Max Planck Institute for Physics, Munich, Germany}
\author[Goe]{Kevin Kr\"oninger\corref{cor1}}
\ead{kevin.kroeninger@phys.uni-goettingen.de}
\cortext[cor1]{Corresponding author}
\address[Goe]{University of G\"ottingen, G\"ottingen, Germany}

\begin{abstract}
We describe the development of a new toolkit for data analysis.  The
analysis package is based on Bayes' Theorem, and is realized with the
use of Markov Chain Monte Carlo. This gives access to the full
posterior probability distribution. Parameter estimation, limit
setting and uncertainty propagation are implemented in a
straightforward manner.  A goodness-of-fit criterion is presented
which is intuitive and of great practical use.\end{abstract}


\end{frontmatter}

\section{Introduction}
\begin{figure}[htbp] 
\centering
\includegraphics[width=0.8\textwidth]{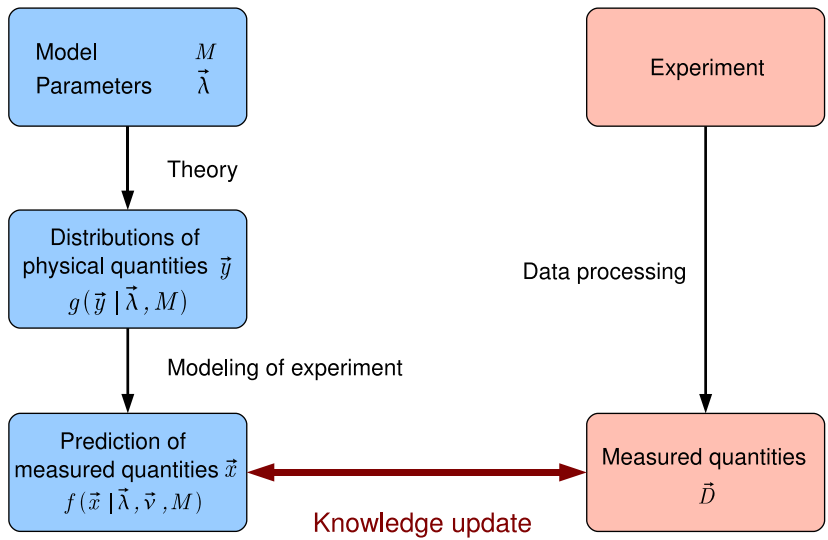} 
\caption{Paradigm for data analysis.  Knowledge is gained from a
comparison of model predictions with data.  Intermediate steps may be
necessary, e.g., to model experimental conditions.}
\label{fig:scheme}
\end{figure}

The goal of data analysis is to compare model predictions with data,
and draw conclusions either on the validity of the model as a
representation of the data, or on the possible values for parameters
within the context of a model. The paradigm for our data analysis
package is shown in Fig.~\ref{fig:scheme}.  Here, the model $M$ can
range from a model purporting to represent nature (e.g., the Standard
Model in particle physics) to a simple parametrization of data useful
for making predictions or for summarizing data.

\subsection{Modeling}
The theory or model can be used to provide {\it direct probabilities};
i.e., relative frequencies of possible outcomes of the results were
one to reproduce the experiment many times under identical conditions.
This is possible because the model is a mathematical construction
which allows the calculation (or simulation) of frequencies of
outcomes.  The predictions from the model cannot usually be directly
compared to experimental results.  An additional step is needed,
either to modify the predictions to allow for the experimental
effects, or to undo the experimental effects from the data.
Obviously, an accurate description of the experimental effects is
necessary to produce reliable conclusions.

The function $g(\vec{y}|\vec{\lambda},M)$ gives the relative frequency
of getting result $\vec{y}$ assuming the model $M$ and parameters
$\vec{\lambda}$. It should satisfy:
\begin{equation}
g(\vec{y}|\vec{\lambda},M)\geq 0
\end{equation}
and
\begin{equation}
\sum_i g(y_i|\vec{\lambda},M) = 1 \;\;\;\; {\rm or} \;\;\;\;\;
\int g(\vec{y}|\vec{\lambda},M) \, d\vec{y} = 1
\end{equation}

\noindent depending on whether discrete or continuous values are
measured.  In the following, we will write formulae for the continuous
case; the modification for the discrete case will be clear.  Note that
the normalization requirement is often discarded when only relative
probabilities of outcomes are needed.

The modeling of the experiment will usually add extra parameters and
assumptions.  We will use the symbol $\vec{\nu}$ to represent these
additional (nuisance) parameters. There could also be additional
information not included explicitly in the model which could limit
values of the model parameters.  This information is often denoted
with an additional symbol, e.g., $I$, and we would then have
$f(\vec{x}|\vec{\lambda},\vec{\nu},M,I)$ for the frequency
distribution of observable quantities $\vec{x}$.  We will drop the $I$
in the following - it is assumed everywhere that all available
information is used in the probability distributions.

As an example, one can consider the radioactive decay of an unstable
nucleus.  The model is an exponential decay time distribution,
\mbox{$P(t|\tau)=\frac{1}{\tau}e^{-t/\tau}$}, with the {\it lifetime}
parameter $\tau$.  Then
$$
g(\{t_i\}|\tau)=\prod_i   P(t_i|\tau)= \prod_i \frac{1}{\tau}e^{-t_i/\tau} \, .
$$

We can imagine that we are measuring a sample of nuclei, and that our
detector is not able to distinguish two decays which occur too closely
in time.  Also, the decay times will be measured with some precision
which could be modeled with a Gaussian smearing with width $\sigma_t$.
This measurement resolution may not be known precisely, and could be
considered a nuisance parameter.  The probability density for a given
set of measured times $f(\{t_j^{\rm measured}\}|\tau,\sigma_t)$ could
then be determined from a simulation which includes the {\it dead
time} and measurement resolution.

In general, the judgment on the validity of a model and the extraction of values of the parameters within the model will be based on a comparison of the data, $\vec{D}$, with $f(\vec{x}|\vec{\lambda},\vec{\nu},M)$.

\subsection{Formulation of the Learning Rule}
The probability of a model, $M$, will be quantified as $P(M)$, with
\begin{equation}
\label{eq:norm}
0 \leq  P(M)  \leq 1 
\end{equation}
while the probability density of the parameters are typically continuous functions.  The parameters from the modeling of the experimental conditions are not correlated to the parameters of the physical model so that
\begin{equation}
P(\vec{\lambda},\vec{\nu}|M)=P(\vec{\lambda}|M)P(\vec{\nu}) \;\; .
\end{equation}
The probability densities satisfy
\begin{eqnarray}
P(\vec{\lambda}|M) & \ge & 0 \\
\int P(\vec{\lambda}|M) \, d\vec{\lambda} &=&1\\
\end{eqnarray}
and similarly
\begin{eqnarray}
P(\vec{\nu}) & \ge & 0 \\
\int P(\vec{\nu}) \, d\vec{\nu} &=&1 \;\; .
\end{eqnarray}

In the Bayesian approach, the quantities $P(M)$ and
$P(\vec{\lambda}|M)$ are treated as probabilities (probability
densities), although they are not in any sense frequency distributions
and are more accurately described as {\it
degrees-of-belief}~\cite{ref:dagostini}.  This {\it degree-of-belief}
is updated by comparing data with the predictions of the models. The
parameter and model {\it degrees-of-belief} are the interesting
quantities since they contain our knowledge about nature.  The purpose
of performing experiments is then to modify our {\it
degree-of-belief}.  $P(M)=1$ represents complete certainty of $M$
being correct, and $P(M)=0$ represents complete certainty that $M$
is false.

The procedure for learning from experiment is:
\begin{equation}
\label{eq:Bayes}
P_{i+1}(\vec{\lambda},\vec{\nu},M|\vec{D}) \propto f(\vec{x}=\vec{D}|\vec{\lambda},\vec{\nu},M) P_{i}(\vec{\lambda},\vec{\nu},M) \, , 
\end{equation}
where the index on $P$ represents a {\it state-of-knowledge}. 

In order to satisfy our normalization requirement 
\begin{equation}
\sum_M \int P(\vec{\lambda},\vec{\nu},M) \, d\vec{\lambda}\, d\vec{\nu}=\sum_M P(M)\left[\int P(\vec{\lambda}|M) \, d\vec{\lambda}\int P(\vec{\nu}) \, d\vec{\nu} \right]=1
\end{equation}
we have
\begin{equation}
\label{eq:learn}
P_{i+1}(\vec{\lambda},\vec{\nu},M|\vec{D}) =\frac{f(\vec{x}=\vec{D}|\vec{\lambda},\vec{\nu},M) P_{i}(\vec{\lambda},\vec{\nu},M)}
{\sum_M \int f(\vec{x}=\vec{D}|\vec{\lambda},\vec{\nu},M) P_{i}(\vec{\lambda},\vec{\nu},M) \, d\vec{\lambda} \, d\vec{\nu}} \;\; .
\end{equation}
We usually just write $P_i=P_0$, and this quantity is called the
{\it prior}. It contains all information we may have on the model and
parameter values {\it before} the experiment is performed.  The
posterior probability density function, $P_{i+1}$, is usually written
simply as $P$. It describes the state of knowledge {\it after} the
experiment is analyzed.

For a given model, $M$, $f$ is a function of the model parameters, the
experimental parameters, and the possible outcomes $\vec{x}$. When $f$
is viewed as a function of $(\vec{\lambda},\vec{\nu})$ for fixed
$\vec{x}=\vec{D}$, it is known as the likelihood.  In our formulation,
$f$ is the relative frequency of a particular result $\vec{x}=\vec{D}$
from our modeling.  If $f$ is normalized, we can write 
\begin{equation}
P(\vec{D}|\vec{\lambda},\vec{\nu},M) =f(\vec{x}=\vec{D}|\vec{\lambda},\vec{\nu},M) \;\; .
\end{equation}

The denominator in Eq.~(\ref{eq:learn}) is the probability to get the
data, $\vec{D}$, assuming the models $M$ and the description of the
experimental conditions describe all possible outcomes, and can be
written as $P(\vec{D})$.  Using this notation, we then recover the
classic equation due to Bayes:
\begin{equation}
P(\vec{\lambda},\vec{\nu},M|\vec{D}) =\frac{P(\vec{D}|\vec{\lambda},\vec{\nu},M) P(\vec{\lambda},\vec{\nu},M)}
{P(\vec{D})} \; .
\end{equation}

The scheme for updating knowledge, as written down here, is general
and straightforward.  There can be difficulties in implementation when
dealing with multi-dimensional spaces, and the toolkit presented here
is designed to help the user solve these technical issues.  It is up
to the user to carefully define the function $f$ as well as the
priors.  The precision and accuracy of the results will depend
primarily on the quality of the inputs. Given that these functions are
well-defined, the {\sc BAT} program will then be useful as a tool for
model testing and parameter estimation.

In the following, we start with a discussion on how to perform
parameter estimation with the model fixed.  This is followed with a
discussion of goodness-of-fit criteria, and the closely related topic
of model comparison. Discovery criteria are discussed in this context.
Setting limits on parameters while keeping the model fixed is briefly
discussed. We then switch to the practical realization of this
analysis framework in terms of Markov Chains.  A brief review of
Markov Chains is given, followed by a discussion of the implementation
in {\sc BAT}. We then give examples of parameter estimation and model
testing to make the ideas more concrete.

\pagebreak 

\section{Parameter Estimation}
\label{section:parameter}
Parameter estimation is performed while keeping the model fixed.  In
this case, we write
\begin{equation}
\label{eq:posterior}
P(\vec{\lambda},\vec{\nu}|\vec{D},M) =\frac{P(\vec{x}=\vec{D}|\vec{\lambda},\vec{\nu},M) P_{0}(\vec{\lambda},\vec{\nu}|M)}
{\int P(\vec{x}=\vec{D}|\vec{\lambda},\vec{\nu},M) P_{0}(\vec{\lambda},\vec{\nu}|M) \, d\vec{\lambda} \, d\vec{\nu}} \;\; .
\end{equation}
The output of the evaluation is a normalized probability density for the parameters, including all correlations.  This output can be used to give best-fit values, probability intervals for the parameters, etc.

The function, $P(\vec{x}=\vec{D}|\vec{\lambda},\vec{\nu},M)$, which
gives the probability density at $\vec{x}=\vec{D}$ given the model $M$
and the parameters $(\vec{\lambda},\vec{\nu})$ must be defined by the
user, and must return reasonable values for all possible data results.
This likelihood function contains the information both from the theory
input as well as the modeling of the experimental conditions.  It
must be defined case-by-case\footnote{Some applications, such as
  histogram fitting using products of Poisson distributions or curve
  fitting using products of Gaussian distributions can be performed in
  {\sc BAT} with minimal effort.}.  Similarly, the prior
probabilities, $P_{0}(\vec{\lambda},\vec{\nu}|M)$, must be defined by
the user.  These prior probability functions define the range over
which the parameters can vary, and any preconceptions concerning their
possible values.  Note that if these priors are set to a constant (not
always applicable!), then the parameter estimation using the mode of
the posterior probability density function ({\it pdf}) is equivalent
to a maximum likelihood estimation.  If in addition the fluctuations
of the data from the model predictions are assumed to follow Gaussian
distributions, then the mode finding reduces to a $\chi^2$
minimization.  The formalism therefore contains these commonly used
fitting approaches.  It is however completely general and is not
limited by these conditions.  In all cases, the full posterior {\it
  pdf} is available, thus allowing the user to study correlations
between parameters, to perform the propagation of uncertainties
without approximations, and to define uncertainty bands or limits as
desired.  The usefulness of this information is demonstrated in the
examples section.

When working with the posterior {\it pdf},
$P(\vec{\lambda},\vec{\nu}|\vec{D},M)$, it is often the case that one
is interested not in the full {\it pdf}, but in the probability distribution
for only one, or a few, parameters.  These distributions are
determined via marginalization.  For example, the probability
distribution for parameter $\lambda_i$ is:
\begin{equation}
\label{eq:marginalization}
P(\lambda_i|\vec{D},M) = \int P(\vec{\lambda},\vec{\nu}|\vec{D},M) \, d\vec{\lambda}_{j\neq i} \, d\vec{\nu} \; .
\end{equation}
Note that the parameter values which maximize the full posterior {\it pdf} usually do not coincide with the values which maximize the marginalized distributions.

\subsection{Use of the Posterior Probability Distribution}
The posterior {\it pdf}, $P(\vec{\lambda},\vec{\nu}|\vec{D},M)$, can be used
to evaluate any desired quantity related to the parameters.  For
example, commonly used quantities are:

\paragraph{Mean of $\lambda_i$} 
$$<\lambda_i> = \int P(\lambda_i|\vec{D},M) \lambda_i \, d\lambda_{i} \;\; .$$
\paragraph{Median of $\lambda_i$} 
The value of $\lambda_i$ such that 50\% of the probabilty is below
this value
$$ \int_{\lambda_{\rm min}}^{\lambda_{\rm med}} P(\lambda_i|\vec{D},M) \, d\lambda_{i}=0.5 \, , $$ 
where $\lambda_{\rm min}$ is the minimum possible value for parameter
$\lambda_i$.  The desired value is $\lambda_{\rm med}$.
\paragraph{Mode of $\lambda_i$} 
The value of $\lambda_{i}$ which maximizes the marginalized posterior
{\it pdf}
$${\rm argmax}\left[P(\lambda_i|\vec{D},M) \right] \;\; .$$
Note that the mode can be evaluated for any number of parameters with
the rest marginalized.  The most common uses are the mode of the full
{\it pdf} and modes evaluated for one parameter at a time.
\paragraph{rms} 
The root-mean-square is defined as usual
$$rms_i = \sqrt{ \left[ \int P(\lambda_i|\vec{D},M) \lambda_i^2 \, d\lambda_{i} - \left(\int P(\lambda_i|\vec{D},M) \lambda_i \, d\lambda_{i}\right)^2 \right]} \;\; .$$
\paragraph{Central Interval} 
The $(1-2\alpha)$ central interval is defined such that a fraction
$\alpha$ of the probability is contained on either side of the
interval
$$ \alpha=\int_{\lambda_{\rm min}}^{\lambda_{\rm lower}} P(\lambda_i|\vec{D},M) \, d\lambda_{i}=
\int_{\lambda_{\rm upper}}^{\lambda_{\rm max}} P(\lambda_i|\vec{D},M)
\, d\lambda_{i} \, , $$ 
where the desired interval is $[\lambda_{\rm lower},\lambda_{\rm upper}]$.
The minimum and maximum allowed values of the parameter are
$\lambda_{\rm min}$ and $\lambda_{\rm max}$, respectively.
\paragraph{Smallest Interval} 
The smallest set of $\vec{\lambda}$ containing $\alpha$ of the probability. The set satisfies $P(\vec{\lambda}|\vec{D},M)>P_{\mathrm{min}}$, where $P_{\mathrm{min}}$ is
defined as $$
\int_{P_{\mathrm{min}}}^{P(\vec{\lambda}^{*}|\vec{D},M)} p(P(\vec{\lambda}|\vec{D},M)) \, dP(\vec{\lambda}|\vec{D},M)=\alpha \, ,
$$ 
where $\vec{\lambda}^{*}$ maximizes the full posterior {\it pdf}. The
expression $p(P(\vec{\lambda}|\vec{D},M))$ is the probability density
of the posterior {\it pdf}.
\paragraph{Correlation} 
The correlation coefficient between two parameters, $\lambda_i$,
$\lambda_j$, can easily be evaluated
\begin{eqnarray*}
\rho_{ij} & = & \frac{<\lambda_i\lambda_j>-<\lambda_i><\lambda_j>}{rms_i \, rms_j} \\
        & = & \frac{\int \int P(\lambda_i,\lambda_j|\vec{D},M) \lambda_i \lambda_j \, d\lambda_{i} \, d\lambda_j
        -\int P(\lambda_i|\vec{D},M) \lambda_i \, d\lambda_{i}\int P(\lambda_j|\vec{D},M) \lambda_j \, d\lambda_{j}}
        {rms_i \, rms_j} \, .
\end{eqnarray*}        

As should be clear, the posterior {\it pdf} contains all the information one
could wish for, and it is just a matter of defining which quantities
are of interest.  A standard output of an analysis could be:
\begin{itemize}
\item the modes of the parameters from the full posterior {\it pdf};
\item the mean and mode of each parameter;
\item the rms of each parameter;
\item the correlation coefficient between each pair of parameters;
\item the central 68\% and 90\% probability intervals and narrowest
  probability intervals for each parameter.
\end{itemize}

\subsection{Using the Posterior for Uncertainty Propagation}
Having full access to the posterior {\it pdf} allows for the evaluation of
any function of the parameters, and the evaluation of the probability
distribution of that function.  In contrast to standard techniques
used for propagation of uncertainties, there is no need here for any
approximations.  As an example, consider a function $y$ of the
variable $x$ which depends on the parameters $\vec{\lambda}$, such as
a linear function, $y=mx+b$, with $\vec{\lambda}=(m,b)$. To evaluate
the probability density distribution for $y$ at any $x$, we have
\begin{equation}
P(y(x)) =  \int P(m,b)\delta(y=mx+b)\, dm \, db \; .
\end{equation}
This can then be used to find any of the quantities of interest regarding the function $y$ at any value of $x$, such as for example the central 68\% interval:
$$ 0.16=\int_{y_{\rm min}}^{y_{\rm lower}} P(y(x)) \, dy=\int_{y_{\rm upper}}^{y_{\rm max}} P(y(x)) \, dy $$
with the central 68\% interval given by $(y_{\rm lower},y_{\rm upper})$.

\section{Model Validity}
\label{sec:model}
\subsection{Definition of $p$-value}
Model testing in a strictly Bayesian approach is problematic since
there is often no way to define all possible models, and the results
depend critically on the choice of priors.  However, having a number
representing how well the model fits the available data is
important. In the {\sc BAT} toolkit, a $p$-value is defined which
gives a goodness-of-fit criterion based on the likelihood of the data
in the model under consideration using the parameters defined at the
mode of the posterior. We define the following quantities:

\begin{eqnarray}
\label{eq:pvalue}
f^*(\vec{x}) & = & P(\vec{x}|\vec{\lambda}^*,\vec{\nu}^*,M) \\
f^D   & = & P(\vec{D}|\vec{\lambda}^*,\vec{\nu}^*,M) =  f^*(\vec{x}=\vec{D}) \, , 
\end{eqnarray}
where $(\vec{\lambda}^*,\vec{\nu}^*)$ is the set of parameter values
at the mode of the full {\it pdf}.  We then define the following
quantity to evaluate the validity of a model, $M$:
\begin{equation}
p=\frac{\int_{f^*(\vec{x})<f^D} f^*(\vec{x}) \, d\vec{x}}
{\int f^*(\vec{x}) \, d\vec{x}} \;\; .
\end{equation}

The quantity $p$ is the {\it tail-area} probability to have found a
result with $f^*<f^D$, assuming that the model $M$ and the parameters
$(\vec{\lambda}^*,\vec{\nu}^*)$ are valid\footnote{A different
  definition used in Bayesian data analysis does not fix the parameter
  value to the mode, but rather weights them with the posterior
  {\it pdf}~\cite{ref:Gelman}.}.  This quantity is analogous to the
well-known $\chi^2$ probability.  Note that before the experiment is
performed, $p$ can be viewed as a random variable.  The definition of
$p$ can be rewritten as:
\begin{equation}
p(\vec{D}) = \int_0^{f^D} P(f^*(\vec{x})) \, df^*(\vec{x})
\end{equation}
so that $p$ is just the value of the cumulative distribution function
for $P(f^*(\vec{x}))$ evaluated at $\vec{x}=\vec{D}$.  This implies
that, assuming the modeling is valid, $p$ will have a flat probability
distribution between $[0,1]$, and is therefore well-suited as a
goodness-of-fit quantity.  It is the probability that the likelihood
could have been lower than the one observed in the data, assuming the
model and parameter values are correct.  If the model does not give a
good representation of the data, then $p$ will be a small number.
This definition is of course only valid if the parameters are not
adjusted with the data at hand.  I.e., the argument given is strictly
only valid for the model with the current values of the parameters
compared to a future data set.  Since the existing data is used to
modify the parameter values, the extracted $p$-value will be biased to
higher values.  The amount of bias will depend on many aspects,
including the number of data, the number of parameters, and the
priors.  The user will need to keep this in mind when using the
$p$-value.  However, we feel that the $p$-value is nevertheless a
useful quantity for evaluating how well a model represents data.

\subsection{Model Comparisons and Discovery Criteria}
Model comparison and deciding on a discovery are related topics. Model
comparison can be performed either via the $p$-value described above,
or via a probability calculated from the posterior {\it pdf}.  There are
different possibilities for formulating the discovery process.  One
possibility is to perform hypothesis testing, and to declare a
discovery if the probability of the hypothesis `the standard physics
model contains all processes' is small.  Another is to calculate the
$p$-value for the `standard model' to explain the observations, and
compare this to the $p$-value of a model including new physics.

\paragraph{Comparison via $p$-values}
Model comparison is easily performed by comparing the $p$-values for
different models.  Large $p$-values imply that the model is a good
representation of the data.  If there are several models available,
then the one with the largest $p$-value gave the best representation,
although all models with reasonable $p$-values (e.g., larger than
$0.1$) should be considered to give good fits.  Note that implementing
Occam's razor is straightforward - one would choose the simplest model
which gives a good $p$-value.  If the $p$-value for the standard
physics model is very small and the $p$-value for a model containing
the new physics is reasonably large, then a discovery could be
claimed.

\paragraph{Comparison via absolute probability}
Models can also be compared by calculating the absolute probability of each model:
\begin{eqnarray}
\label{eq:probab}
P(M|\vec{D})&=&\int P(M,\vec{\lambda},\vec{\nu}|\vec{D}) \, d\vec{\lambda} \, d\vec{\nu} \\
           &=&\frac{\int P(\vec{D}|\vec{\lambda},\vec{\nu},M)P_0(\vec{\lambda},\vec{\nu},M) \, d\vec{\lambda} \, d\vec{\nu}}{\sum_M \int P(\vec{D}|\vec{\lambda},\vec{\nu},M)P_0(\vec{\lambda},\vec{\nu},M) \, d\vec{\lambda} \, d\vec{\nu}} \\
           &=&\frac{\int P(\vec{D}|\vec{\lambda},\vec{\nu},M)P_0(\vec{\lambda},\vec{\nu}|M) \, d\vec{\lambda} \, d\vec{\nu} \; P_0(M)}{\sum_M \int P(\vec{D}|\vec{\lambda},\vec{\nu},M)P_0(\vec{\lambda},\vec{\nu}) \, d\vec{\lambda} \, d\vec{\nu}\; P_0(M)} \, .
\end{eqnarray}
This approach has the advantage of being a true probability ({\it
  degree-of-belief}), but it requires the specification of a full set
of models giving
\begin{equation}
\sum_M P_0(M)=1 
\end{equation}
and is often very sensitive to the definitions of the priors.  It has
also a technical disadvantage in that the denominator must be
calculated (it is not necessary for the other quantities discussed
thus far).  An example of this approach is given in~\cite{ref:GERDA},
where the search for neutrinoless double beta-decay is described.
There, only two possibilities are allowed - there are only background
processes, or, there are background processes and additionally a
neutrinoless double beta-decay signal.  It is possible to formulate
the priors and likelihoods for each of these cases, and a discovery
criterion can be given as $P({\rm background \; only}|\vec{D})<P_{\rm
threshold}$, where $P_{\rm threshold}$ is a small value.

\section{Setting limits on Parameters}
Setting limits on parameters is normally straightforward, and just
requires integration of the posterior {\it pdf}.  For example, setting a
90\% upper limit on parameter $\lambda_i$ would require solving
$$0.9=\int_{\lambda_{\rm min}}^{\lambda_{\rm upper}} P(\lambda_i|\vec{D},M) \,
d\lambda_i$$ for $\lambda_{\rm upper}$.  However, there are situations in
which the posterior cannot be evaluated because there is great
uncertainty on the form of the prior, and the posterior {\it pdf} depends
strongly on the form chosen.  This type of situation can arise when
probability distributions for parameters of a new, unproven, model are
desired, and the goal is to rule out certain parameter ranges for the
model.  One possibility in this case is to use the ratio of
likelihoods as proposed by d'Agostini~\cite{ref:dagostini}:
$$R(\vec{\lambda}|\vec{D},M)=\frac{P(\vec{D}|\vec{\lambda},M)}{P(\vec{D}|S)} \, ,$$
where $S$ represents the standard model.  One can then choose by
convention to say that parameter values yielding $R<R_{\rm cut}$ are
ruled-out (better, disfavored).  A possible choice for $R_{cut}$ is
$0.1$.  This does not mean that the parameter values giving small $R$
are ruled out with a certain probability - there is not enough
information to say this - but the {\it degree-of-belief} in these
values is certainly low.

\pagebreak 

\section{Markov Chain Monte Carlo}
The posterior {\it pdf}, Eq.~(\ref{eq:Bayes}), is determined using a Markov
Chain Monte Carlo ({\it MCMC})~\cite{ref:Markov}.  We give a brief review of
Markov Chains, and the particular algorithm used to realize the chain.

\subsection{Markov Chains}

\begin{figure}[htbp] 
   \centering
   \includegraphics[width=0.8\textwidth]{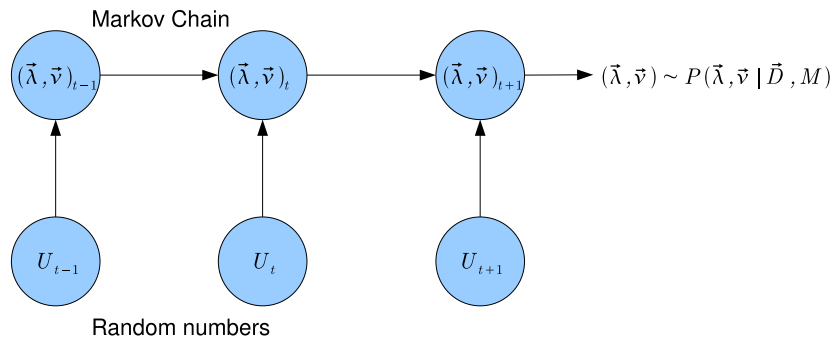} 
   \caption{Scheme for producing a Markov Chain with stationary
     distribution $P(\vec{\lambda},\vec{\nu}|\vec{D},M)$. Independent,
     identically distributed random numbers are used to create random
     numbers, or vectors of numbers, according to any desired
     distribution.}
   \label{fig:MCMC}
\end{figure}

Markov Chains are sequences of random numbers (or vectors of numbers),
$X_t$, which have a well-defined limiting distribution, $\pi(x)$.  The
fundamental property of a Markov Chain is that the probability
distribution for the next element in the sequence, $X_{t+1}$, depends
only on the current state, and not on any previous history.  A Markov
Chain is completely defined by the one-step probability transition
matrix, $P(X_{t+1}=y|X_t=x)$.  Under certain conditions (recurrence,
irreducibility, aperiodicity), it can be proven that the chain is
ergodic; i.e., that the limiting probability to be in a given state
does not depend on the starting point of the chain.  The probability
to be in a given state is then stationary.  An {\it MCMC} is a method
producing an ergodic Markov Chain which stationary distribution is the
distribution of interest.  In our case, we produce a Markov Chain
where the stationary probability density is the desired posterior {\it
pdf}; i.e., $\pi=P(\vec{\lambda},\vec{\nu}|\vec{D},M)$.  A graphical
description is given in Fig.~\ref{fig:MCMC}.

The original and most popular algorithm which achieves this is the Metropolis algorithm~\cite{ref:Metropolis}.  

\subsection{Metropolis Algorithm}
 The algorithm works as follows:
\begin{enumerate}
\item Given that the system is in state $X_t=\vec{x}$, a new proposed
  state, $\vec{y}$, is generated according to a symmetric proposal
  function, $g(\vec{y},\vec{x})$.  In our application, a state is a
  particular set of parameter values.
\item The quantity $$r=\frac{\pi(\vec{y})}{\pi(\vec{x})}$$ is then
  calculated, and compared to a random number $U$, generated flat
  between $[0,1]$.  If $U<r$, then we set $X_{t+1}=\vec{y}$, else, we
  take $X_{t+1}=\vec{x}$.
\end{enumerate}
It is possible to show that, given a reasonable proposal function $g$,
this algorithm satisfies the conditions of an {\it MCMC}, and that the
limiting distribution is $\pi(\vec{x})$.  This then allows for the
production of states distributed according to the desired
distribution.  In particular, this allows for the generation of
randomly distributed system states according to complicated
probability density functions which have no analytic form.  All that
is required is that $\pi(\vec{x})$ can somehow be calculated.

\section{Implementation}

The requirements for a technical implementation of the Bayesian
inference described above are (i) a flexible framework which allows to
formulate the models developed, and (ii) a reliable and fast code for
numerical operations such as optimization, marginalization and
integration. The {\sc BAT} software framework was designed to fulfill
both requirements. Its technical implementation is described in the
following.

\subsection{Framework}

The {\sc BAT} software framework is C++ based code which comes in the
form of a library. It offers classes which are used to phrase the
problems encountered and to perform numerical operations. The
framework is interfaced to software packages such as {\sc
  ROOT}~\cite{ROOT}, {\sc Minuit}~\cite{Minuit}, or the {\sc CUBA}
library~\cite{CUBA}. It offers several output formats, e.g., plain
ASCII files, {\sc ROOT} trees and graphical displays. The {\sc BAT}
library can be linked against in a standalone program or be loaded
into {\sc ROOT}. \\

The formulation of models and their corresponding terms, as, e.g., in
Eq.~(\ref{eq:learn}), is done in the form of methods which belong to
the above mentioned classes. These are defined by the user. For simple
applications such as fitting a histogram or a function with Poissonian
or Gaussian uncertainties, the {\sc BAT} software provides predefined
classes which can be used with minimal programming effort. \\

A detailed description of the class structure and the methods is given
in~\cite{website}.

\subsection{Numerical Implementation}

The numerical operations which need to be performed during the
analysis are optimization, marginalization and integration. Several
algorithms for each operation are either implemented in the current
version of the code or planned for later versions. Emphasis is placed
on the {\it MCMC}.

\paragraph{{\it MCMC}} The posterior probability density
of Eq.~(\ref{eq:posterior}) is sampled using an {\it MCMC}. A {\it
  pre-run} of the {\it MCMC} is performed before the {\it sampling and
  analysis run} in order to ensure convergence and to find reasonable
run parameters. \\

For the {\it pre-run}, several chains are run in parallel with random
starting points in the allowed range. The steps in parameter space are
done consecutively for each parameter and chain. The proposal function
for new steps is chosen flat by default with a range initially set to
the width of the allowed range of the parameter. An iteration is
defined as the set of steps from an update of the first parameter of
the first chain to the last parameter of the last chain. The
efficiency for accepting or rejecting new points is evaluated
separately for each parameter and chain over several iterations
(default: $1,000$). The proposal function ranges are updated every
$1,000$ iterations until an efficiency between 10\% and 50\% is found
for each parameter. The convergence of Markov chains is monitored via
the $r$-value~\cite{rvalue}, which is defined as $r=\sqrt{\hat{V}/W}$,
where $\hat{V}$ is the estimated variance of the target distribution
and $W$ is the mean of variances of all chains. The quantities are
defined as follows:
\begin{eqnarray*}
W & = & \frac{1}{m} \frac{1}{n-1} \sum_{j=1}^{m} \sum_{i=1}^{n} (x_{i} - 
\bar{x}_{j})^{2} \, , \\
\hat{V} & = & (1-\frac{1}{n}) W + \frac{1}{m-1} \sum_{j=1}^{m} (\bar{x}_{j} - \bar{x})^{2} \, , 
\end{eqnarray*}
where $m$ is the number of chains run simultaneously (default: 5), $n$
is the number of elements in one chain for which the variance is
estimated (default: $1,000$ iterations), and $x$ is a quantity of
interest.

Once convergence is reached both $\hat{V}$ and $W$ should be the same,
i.e., $r\approx 1$.  The convergence criterion is set by default to
$(r-1)<0.1$ has to be fulfilled for all parameters simultaneously, as
well as by the posterior {\it pdf}.  The {\it pre-run} is performed
for a minimum number of iterations (default: $500$) and is continued
either until the chains converged and the efficiency of each parameter
is found in the required range, or until a maximum number of
iterations is reached (default: $1,000,000$). No output is produced
during the {\it pre-run}.

The sampling and analysis run is performed for a defined number of
iterations (default: $100,000$) with run parameters found in the
{\it pre-run}. Several operations are performed during the sampling: the
global mode is compared to the current point, histograms are filled
for the marginalized distributions (Eq.~(\ref{eq:marginalization})),
and, if applicable, the probability distribution for the function
being fitted to the data is evaluated. A user interface allows to
perform further operations during the sampling such as evaluating
arbitrary functions of the parameters. \\

\paragraph{Optimization} The {\sc BAT} package extracts the mode as a
standard output. In case this is the only information from the
posterior {\it pdf} of interest, an interface to the {\sc ROOT}
version of {\sc Minuit} can be used. All necessary information, such
as the ranges of the parameters and the function itself, are
automatically transfered to {\sc Minuit}. The {\sc Minuit} run
parameters can be adjusted by the user.

\paragraph{Integration} For model comparisons via the posterior probability of 
the models the denominator of Eq.~(\ref{eq:probab}) has to be
evaluated. If no analytical expression is provided by the user, a
numerical integration over the posterior probability is performed. In
the current version the implemented algorithms are {\it sampled mean}
with and without {\it importance sampling}. Interfaces to tools such
as the CUBA library are available. \\

\paragraph{Evaluation of the validity of a model} The validity of a
model can be evaluated with the prescription given in
section~\ref{sec:model}. In cases where the likelihood is the product
of a finite number of simple expressions, {\sc BAT} can generate
ensemble data sets with the parameters which maximize the posterior
{\it pdf}. Subsequently, the likelihood is calculated for those data
sets and the $p$-value is estimated. In case the likelihood is more
complicated, an external generator has to be used. The ensemble data
sets can be read into {\sc BAT} and used for the analysis.

For the cases where the likelihood is a product of Gaussian or
Poissonian distributions, a fast evaluation of the $p$-value is
possible in {\sc BAT}.

\subsection{Output}

{\sc BAT} provides the following output by default:

\begin{itemize} 

\item a plain ASCII file which summarizes the models and their
  parameters. It displays the results of the analysis such as the
  global mode, the mean and modes of the marginalized distributions,
  or the uncertainties on the parameters;

\item ROOT trees which store the summary information. The information
  of several runs with, e.g., different data sets can be stored and
  compared;

\item ROOT trees containing the Markov Chains. All points of the
  chains can be stored together with an index and the posterior
  {\it pdf} at that point. This allows the chains to be
  analyzed offline;

\item histograms of the marginalized distributions (1D) and histograms
  containing the correlation between two parameters (2D). These are
  stored in the form of ROOT histograms.

\end{itemize}

\section{Examples}
As examples of the use of {\sc BAT}, we consider the analysis of two
similar data sets generated using the same functional form.  In the
first case, the data fluctuations are taken to be Gaussian while in
the second case the fluctuations are generated using a Poisson
distribution. The same models are used to fit to the data in each
case, but the conclusions are quite different, driven in large part by
the different uncertainties assigned to the data points.  We first
discuss the Gaussian fitting example in some length, describing the
fits and the standard {\sc BAT} output.  The wealth of information
which can be extracted from the Markov Chain is apparent already in
these simple examples.  A comparison of the use of the $p$-value
proposed in this paper with the posterior probability of the models is
also discussed.

\subsection{Example: Gaussian Uncertainties}
\begin{figure}[htbp] 
   \centering
   \includegraphics[height=0.45\textwidth]{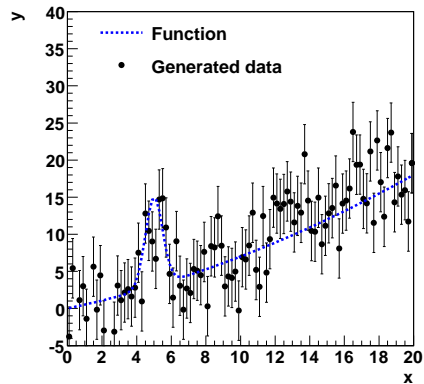}
   \caption{The data set used in the example fits described in the
     text. The function from which the data points were generated is
     shown as a dashed blue line.}  \label{fig:data}
\end{figure}

The data shown in Fig.~\ref{fig:data} were fitted by several models.
The data point $y_i$ at a given $x_{i}$ was generated by sampling from
a Gaussian distribution with mean $f(x_{i})$ and fixed standard
deviation of $s=4$ units, where
\begin{equation}
\label{eq:Data}
f(x_{i}) = A + B \, x_{i} + C \, x_{i}^{2} + \frac{D}{\sigma \sqrt{2 \pi}} e^{- \frac{(x_{i}-\mu)^{2}}{2\sigma^{2}}} \, , 
\end{equation}
and ($A=0$, $B=0.5$, $C=0.02$, $D=15$, $\sigma=0.5$, $\mu=5.0$).  The
data are plotted in the usual way, with the bar on each data point
indicating the size of the data uncertainty (in this case the value
used to generate the data).

\pagebreak 

The data were fitted using the following models:
\begin{enumerate}
\item second order polynomial;
\item constant plus Gaussian;
\item linear plus Gaussian;
\item quadratic plus Gaussian.
\end{enumerate}

The likelihood function was taken to be
$$
P(\vec{D}|\vec{\lambda},M)=\prod_{i} \frac{1}{\sqrt{2\pi}s}e^{-\frac{(y_i-f(x_{i}))^2}{2s^2}} \, , 
$$
where $f(x_{i})$ is evaluated using the current values of the
parameters $\vec{\lambda}$ in the model. Table~\ref{tab:models}
summarizes the parameters available in each model and the range over
which they are allowed to vary. In all models, flat priors were
assumed for all parameters. An example with non-flat priors is
discussed in a later section.

\begin{sidewaystable}[htdp]
\begin{center}
\begin{tabular}{clccccccc}
\hline
& Model & Par & Min & Max & Global & Mean/  & Central $68\%$  & $p$-value \\
&  & & & & mode &  Limit (95\%) & interval &  \\
\hline
I.  & $A_{\rm I} + B_{\rm I} \, x_{i} + C_{\rm I} \, x_{i}^{2}$  & $A_{\rm I}$ & \phantom{-}0.0 & $\phantom{00}5.0$ & $0.5\cdot10^{-3}$ & $<1.7$ & - & $0.154$ \\
    & & $B_{\rm I}$ & \phantom{-}0.0 & \phantom{00}1.2 & \phantom{00}0.77 & \phantom{00}0.65 & \phantom{00}0.48--\phantom{00}0.81 & \\ 
    & & $C_{\rm I}$ & -0.1           & \phantom{00}0.1 & $9.4\cdot10^{-3}$ & $14.6\cdot10^{-3}$ & $(5.1-24.1) \cdot 10^{-3}$  & \\ 
\hline
II. & $A_{\rm II} + \frac{D_{\rm II}}{\sigma_{\rm II} \, \sqrt{2 \pi}} e^{- \frac{(x_{i}-\mu_{\rm II})^{2}}{2\sigma_{\rm II}^{2}}}$ & $A_{\rm II}$ & \phantom{-}0.0 & \phantom{0}10.0 & \phantom{00}3.5\phantom{0} & \phantom{00}3.6\phantom{0} & \phantom{00}3.0\phantom{0}--\phantom{00}4.2\phantom{0} & 0.025 \\ 
 & & $D_{\rm II}$      & \phantom{-}0.0 & 200.0           & 142.9\phantom{0}           & 136.8\phantom{0} & 125.6\phantom{0}--147.9\phantom{0} & \\
 & & $\mu_{\rm II}$    & \phantom{-}2.0 & \phantom{0}18.0 & \phantom{0}17.2\phantom{0} & \phantom{0}17.1\phantom{0} & \phantom{0}16.7\phantom{0}--\phantom{0}17.5\phantom{0} & \\ 
 & & $\sigma_{\rm II}$ & \phantom{-}0.2 & \phantom{00}4.0\phantom{0}& \phantom{00}4.0\phantom{0}\phantom{0} & $>3.5$ & - & \\
\hline
III. & $A_{\rm III} + B_{\rm III} \, x_{i} $ & $A_{\rm III}$ & \phantom{-}0.0 & \phantom{0}10.0 & $0.6\cdot10^{-3}$ & $<1.5$ & - & 0.479 \\ 
     & $+ \frac{D_{\rm III}}{\sigma_{\rm III} \, \sqrt{2 \pi}} e^{- \frac{(x_{i}-\mu_{\rm III})^{2}}{2\sigma_{\rm III}^{2}}}$ & $B_{\rm III}$      & \phantom{-}0.0 & \phantom{00}2.0 & \phantom{00}0.89            & \phantom{00}0.77           & \phantom{00}0.63 -- \phantom{00}0.89 & \\
 & & $D_{\rm III}$      & \phantom{-}0.0 & 200.0           & \phantom{00}9.8\phantom{0}  & \phantom{0}23.6\phantom{0} & \phantom{00}6.6\phantom{0} -- \phantom{00}4.6\phantom{0} & \\
 & & $\mu_{\rm III}$    & \phantom{-}2.0 & \phantom{0}18.0 & \phantom{00}5.2\phantom{0} & \phantom{0}11.9\phantom{0} & \phantom{00}5.1\phantom{0} -- \phantom{0}17.2\phantom{0} & \\ 
 & & $\sigma_{\rm III}$ & \phantom{-}0.2 & \phantom{00}4.0\phantom{0} & \phantom{00}0.5\phantom{0} & \phantom{00}2.1\phantom{0} & \phantom{00}0.5\phantom{0} -- \phantom{00}3.6\phantom{0} & \\
\hline
IV. & $A_{\rm IV} + B_{\rm IV} \, x_{i} + C_{\rm IV} \, x_{i}^{2} $ & $A_{\rm IV}$ & \phantom{-}0.0 & \phantom{0}10.0 & $3.6\cdot10^{-3}$ & $<1.4$ & - & 0.667 \\ 
    & $+ \frac{D_{\rm IV}}{\sigma_{\rm IV} \, \sqrt{2 \pi}} e^{- \frac{(x_{i}-\mu_{\rm IV})^{2}}{2\sigma_{\rm IV}^{2}}}$  & $B_{\rm IV}$      & \phantom{-}0.0 & \phantom{00}2.0 & \phantom{00}0.51           & \phantom{00}0.42   & \phantom{00}0.26--\phantom{00}5.84 & \\
 & & $C_{\rm IV}$      & \phantom{-}0.0 & \phantom{00}0.5 & $25.4\cdot10^{-3}$         & $28.1\cdot10^{-3}$  & $(18.1-3.8)\cdot10^{-3}$ & \\
 & & $D_{\rm IV}$      & \phantom{-}0.0 & 200.0           & \phantom{0}12.56           & \phantom{0}12.8\phantom{0}  & \phantom{00}8.9\phantom{0} -- \phantom{0}16.3\phantom{0} & \\
 & & $\mu_{\rm IV}$    & \phantom{-}2.0 & \phantom{0}18.0 & \phantom{00}5.2\phantom{0} & \phantom{00}5.5\phantom{0}    & \phantom{00}4.9\phantom{0} -- \phantom{00}5.4\phantom{0} & \\ 
 & & $\sigma_{\rm IV}$ & \phantom{-}0.2 & \phantom{00}4.0 & \phantom{00}0.55    & \phantom{00}0.75 & \phantom{00}0.48 -- \phantom{00}0.87 & \\
\hline
\end{tabular}
\end{center}
\caption{Summary of the models fitted to the data, along with the
  ranges allowed for the parameters. The value of the parameter from
  the global mode, the mean (or $95$\% probability limit for a
  parameter if it is at the edge of the allowed range) and the central
  $68$\% probability interval from the marginalized distribution are
  given. The last column gives the $p$-value of the fit.}
\label{tab:models}
\end{sidewaystable}%

\begin{figure}[htbp] 
   \centering
   \includegraphics[width=0.9\textwidth]{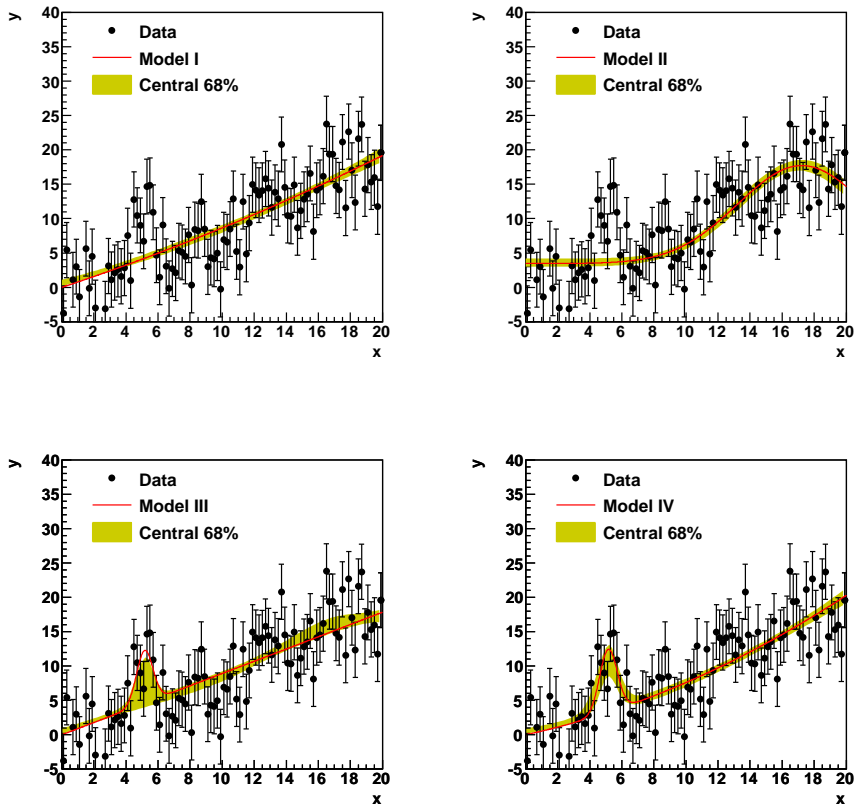}
   \caption{Graphical display of the model fits to the data.  The
     solid lines give the models with the parameters set to the global
     mode values.  The band is calculated as explained in the text.}
  \label{fig:sigma4}
\end{figure}

The fits themselves are shown in Fig.~\ref{fig:sigma4}, while the
results for the mode, the mean and the central 68\% intervals on the
parameters are given in Tab.~\ref{tab:models}. The solid lines in
Fig.~\ref{fig:sigma4} are the functional forms evaluated with the
parameters set to the global mode values. For the calculation of the
uncertainty band, the output of the Markov Chain was used to produce a
distribution of $y$-values from the model at different $x$-values. The
central 68\% probability interval of these $y$-values then defines the
uncertainty band. We note the following:
\begin{itemize}
\item the fit functions all give a reasonable description of the data
  within the constraints allowed by the functional forms;
\item the values of the functions evaluated at the parameter values
  from the global mode can lie outside the central probability
  interval (e.g., the peak region in model~III). 
\end{itemize}

The possibility of having multiple modes in the probability
distributions as well as a different definition of the uncertainty
interval will be discussed later.

\subsubsection{The Posterior {\it pdf}}
\begin{figure}[htbp] 
   \centering
   \includegraphics[width=0.9\textwidth]{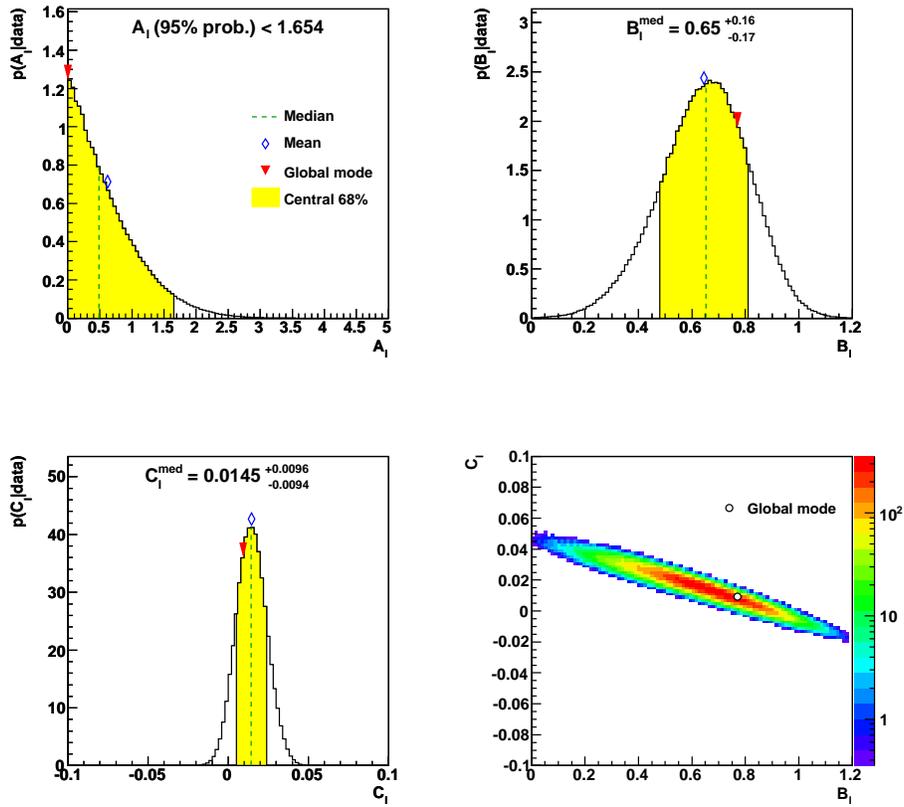}
   \caption{Examples of standard output distributions from the {\sc
   BAT} program. The marginalized {\it pdf}s for each of the
   parameters from model~I are shown as well as the probability
   contours from $P(B_{\rm I},C_{\rm I}|\vec{D})$. Note the log-scale
   in the bottom right plot.}
\label{fig:model1}
\end{figure}

The {\sc BAT} program outputs a file containing sets of parameter
values distributed according to the full posterior {\it pdf}. In
addition, the {\sc BAT} program produces graphical output of the
marginalized posterior {\it pdf}s of all parameters and parameter
pairs. The posterior {\it pdf}s of the parameters for model~I are
shown in Fig.~\ref{fig:model1} as an example of these marginalized
distributions.  For the one dimensional distributions, the dashed line marks the median. 
The mean and global mode are marked by the diamond and the triangle, respectively. 
The band gives the
central $68$\% probability interval.  If the marginalized mode for a
parameter is at one of the limits, the $95$\% probability lowest or
highest interval is shown instead. The median is given
at the top of each plot along with the range allowed from the central
$68$\% probability interval. In the case where a limit is reported,
the upper or lower $95$\% probability value is quoted. For the two
dimensional distribution the {\it pdf} is represented by a color code
and the open circle marks the global mode.

\subsubsection{Multiple Modes}

\begin{figure}[htbp] 
   \centering
   \includegraphics[width=0.9\textwidth]{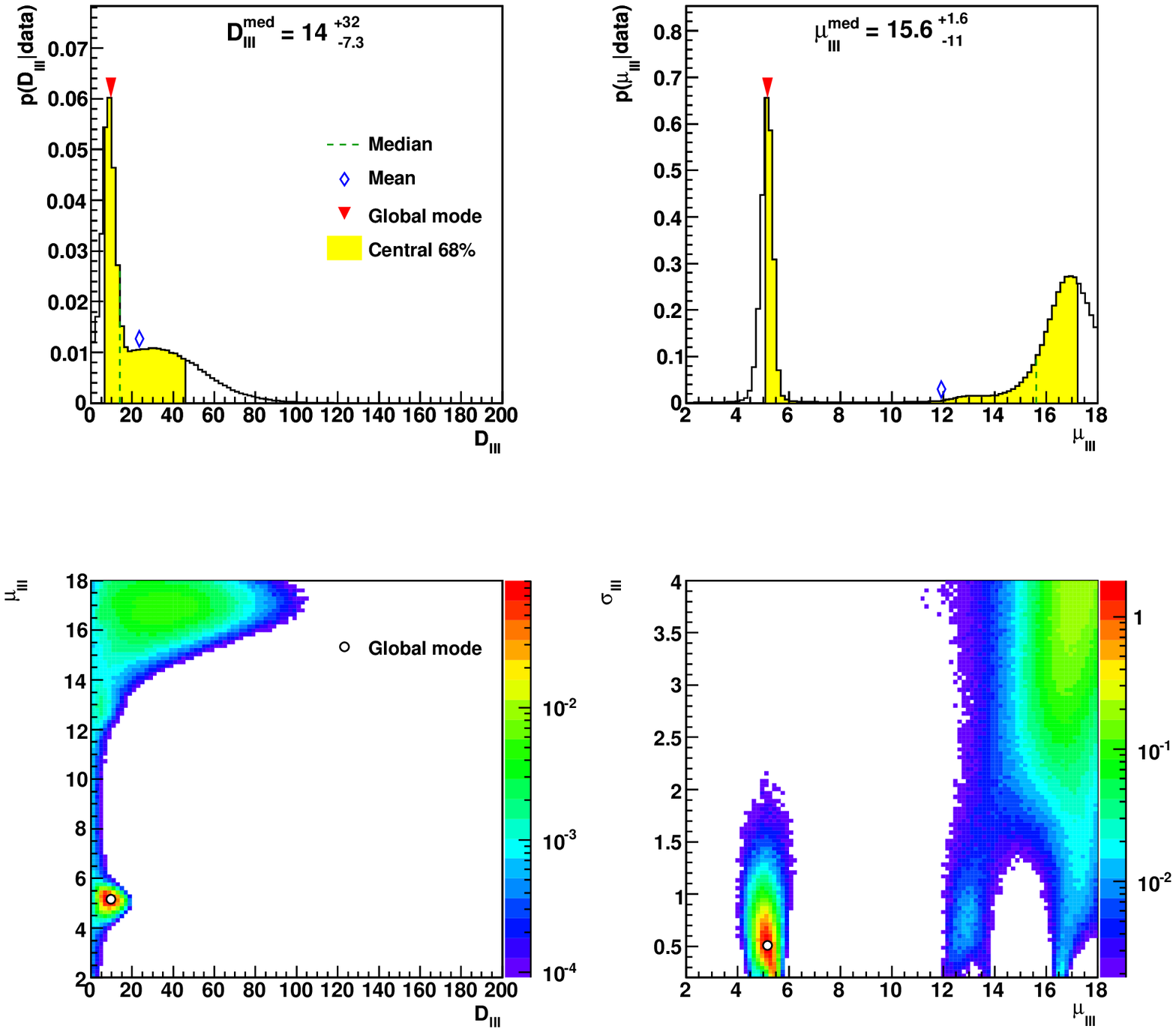}
   \caption{The marginalized {\it pdf}s for some of the parameters
   from model~III. See text for discussion.}  \label{fig:model3}
\end{figure}

It is often the case that there are several modes in the posterior
{\it pdf}. As an example, we look in more detail at the posterior {\it
pdf}s for model~III. Figure~\ref{fig:model3} shows the marginalized
{\it pdf}s for the amplitude, $D_{\rm III}$, and the mean of the
Gaussian peak, $\mu_{\rm III}$, and the two-dimensional probability
contours for the amplitude and mean, and the mean and width of the
Gaussian, $\sigma_{\rm III}$.  As becomes clear from these plots, it
is possible for a large fraction of the probability distribution to be
far from the parameter value of the global mode. The global mode can
in fact lie outside the central 68\% probability interval. In the fit
of model III to our simulated data, two regions for the parameter
$\mu_{\rm III}$ have significant probability. In one case, a low lying
peak is located at $\mu_{\rm III}\approx5$, while in the second case
there is a peak located at $\mu_{\rm III}\approx 17$ (see
Fig.~\ref{fig:model3} upper right).  The peak at $\mu_{\rm
III}\approx5$ corresponds to a narrow Gaussian with small amplitude as
can be seen in the lower left and lower right plot of
Fig.~\ref{fig:model3}. The peak at $\mu_{\rm III}\approx17$ is broad
and has a larger amplitude. The 2D-projections of the posterior {\it
pdf} indicate that there are several regions with significant
probability.

With {\sc BAT} the level of information available allows the user to
see effects such as the multiple modes clearly, and to then react
accordingly.  In the example discussed here, the user may want to
redefine the prior ruling out the possibility of the wide Gaussian and
perform the fit again.

\subsubsection{Definition of Uncertainty Interval}

\begin{figure}[htbp] 
   \centering
   \includegraphics[width=0.9\textwidth]{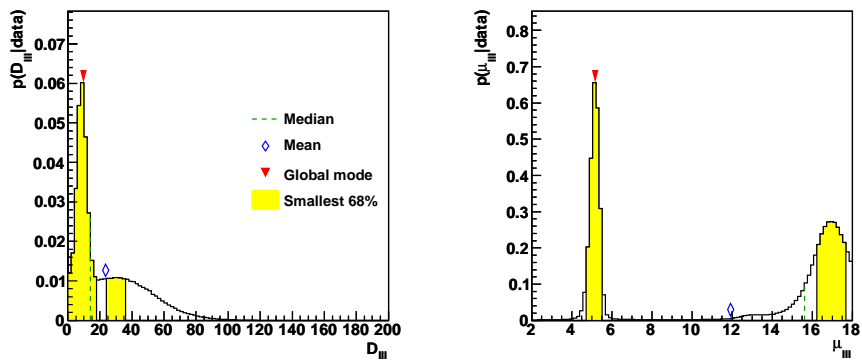}
   \caption{The marginalized {\it pdf}s for parameters $D_{\rm III}$
   and $\mu_{\rm III}$ from model~III using the smallest set of
   intervals containing $68$\% of the probability.}
   \label{fig:model3_si}
\end{figure}

\begin{figure}[htbp] 
   \centering
   \includegraphics[width=0.9\textwidth]{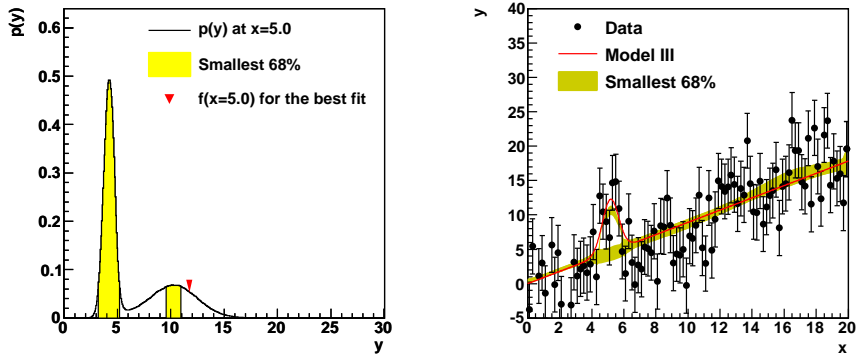}
   \caption{Results from fits of model~III to the data. Left: An
   example for the distribution of $y$-values for a fixed value of
   $x=5.0$. The shaded area is the smallest set of intervals
   containing 68\% of the probability. Right: The uncertainty band
   calculated using this definition.}
\label{fig:twoband}
\end{figure}

The central interval is not always the appropriate choice for the
definition of the uncertainty interval. Another option is to take the
narrowest set of intervals containing $68$\% of the posterior {\it
pdf}. Choosing this definition can produce discontinuous regions in
the parameter space as shown in Fig.~\ref{fig:model3_si}. The
uncertainty band on the value of the fit function at a given value of
$x$ can also be calculated with this approach. As an example,
Fig.~\ref{fig:twoband} (left) shows the distribution of possible
$y$-values for $x=5.0$. The shaded area is the smallest set of
intervals containing $68$\% probability. The resulting uncertainty
band is shown in Fig.~\ref{fig:twoband} (right).

\subsubsection{Model Comparison}
The model comparison can be carried out via the $p$-values given in
Tab.~\ref{tab:models}. The corresponding $p$-values for models I--IV
are 0.154, 0.025, 0.479 and~0.667, respectively. From the $p$-values,
one concludes that models~I, III and~IV are all in good agreement with
the data, whereas model~II is somewhat disfavored.  This is consistent
with the conclusions one would draw from a visual inspection of the
plots in Fig.~\ref{fig:sigma4}. For this example, the $p$-value
provides a useful {\it goodness-of-fit} number summarizing the ability
of a model to represent the data.

It is also possible to calculate directly a posterior probability for
each model to be correct.  In this case, the prior probabilities for
each model were taken to be the same. The posterior probabilities are
then calculated according to Eq.~(\ref{eq:probab}).  The results for
the four models were $0.88$, $7.6\cdot10^{-6}$, $0.12$ and $8.2\cdot
10^{-3}$, respectively.  These values are very sensitive to the range
allowed for the parameter values in the model, and models with more
parameters are automatically disfavored.  Making use of this
probability requires a very considered choice of the priors used for
the models as well as the priors on the individual parameters within a
model.

\subsection{Example: Poissonian Uncertainties}
\begin{figure}[htbp] 
   \centering
   \includegraphics[width=0.9\textwidth]{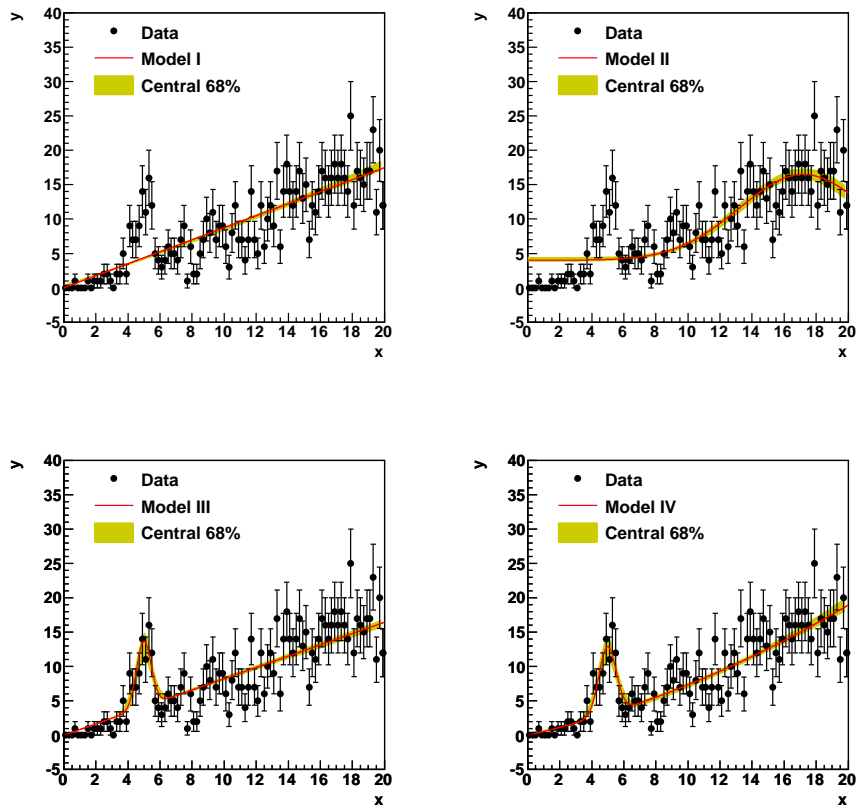}
   \caption{The data set as well as the fitted functions for the
   example with Poissonian fluctuations. Upper left, model~I; upper
   right, model~II; lower left, model~III; lower right, model~IV.  The
   shaded band shows the uncertainty on the fit functions from the
   central 68\% interval.}  
\label{fig:Poisson}
\end{figure}

In this example, the data were generated assuming the expression in
Eq.~(\ref{eq:Data}) is the expectation value for a Poisson
distribution in bin $i$. I.e., the number of events in the $i$th bin,
$n_{i}$, was simulated using a Poisson distributed number with mean
$f(x_{i})$. The same models were fitted to the data, and the parameter
ranges were those specified in Tab.~\ref{tab:models}. The data which
were fitted are shown in Fig.~\ref{fig:Poisson}, together with the
models evaluated using the parameter values from the global mode of
the fits.

\pagebreak 

This example represents a typical histogram fitting application. The
data in bins represent the result of a counting experiment, and the
likelihood function is
$$P(\vec{D}|\vec{\lambda},M)=\prod_{i} \frac{e^{-f(x_{i})} \, f(x_{i})^{n_i}}{n_i!} \, ,$$
where $f(x_{i})$ is the prediction for the mean for bin $i$ for a
given model, $M$.

\begin{figure}[htbp] 
   \centering
   \includegraphics[width=0.9\textwidth]{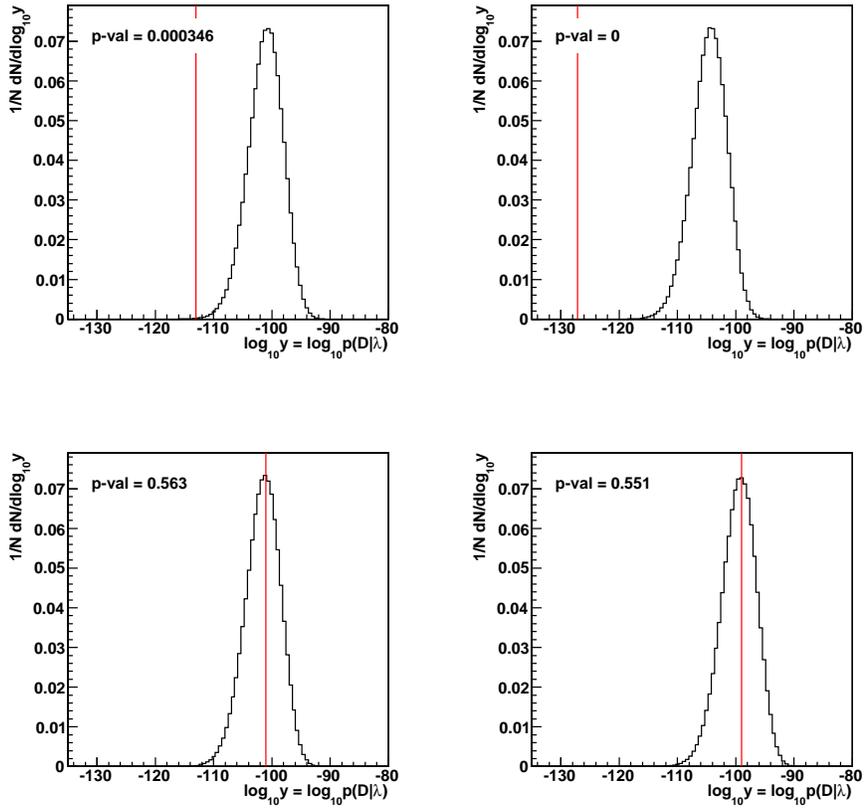}
   \caption{Distribution of $f^{*}$ for models~I--IV (histogram). The
     value of $f^{D}$ is shown as the red vertical line. The $p$-values
     are calculated based on~Eq.~(\ref{eq:pvalue}).}
     \label{fig:pvalues}
\end{figure}

In this case, we find that the models are easily distinguished. There
is a much stronger discrimination between the models since the
uncertainties on the data are considerably smaller.  The distribution
of $f^*$ (see Eq.~(\ref{eq:pvalue})) for $10^{6}$ simulated
experiments for each of the models is shown in Fig.~\ref{fig:pvalues},
along with the value of $f^D$.  The resulting $p$-values are
$0.3\cdot10^{-3}$, $<3\cdot 10^{-6}$, $0.563$ and $0.551$ for
models~I--IV, respectively.  The upper limit on the $p$-value for the
second model results from finding $0$ simulated experiments with such
a low likelihood, and corresponds to a $95$\% upper limit. The
models~I and~II can be clearly distinguished from the models~III
and~IV. It is not possible to distinguish the latter two models based
on the $p$-value, and in practice the model with fewer parameters
would be preferred.

\subsubsection{Non-flat priors} 
As an example of the influence of the prior used for an individual
parameter in the extraction of the posterior {\it pdf}, a different
prior was used for the parameter $\sigma$ in models~II, III and
IV. This restriction could come from a known detector resolution,
e.g., which would limit the observed width of any bumps in the
data. Rather than using a flat prior, the following form was chosen:
$$P_0(\sigma)=\frac{1}{0.3 \sqrt{2\pi}}e^{-\frac{(\sigma-0.7)^2}{0.18}} \, .$$
No significant differences in the marginalized distributions and the
$p$-values were found for models~III and~IV, since the data already
constrain these quite well. However, the solution with a large value
of $\sigma_{\rm II}$ in model~II, discussed above, was now suppressed,
and $f^{D}$ for the fit was reduced.  This clearly illustrates
the need to put the maximum amount of knowledge into the fitting
procedure in order to get out the best possible results.

\pagebreak 

\section{Summary}
We have described the development of a general purpose analysis tool
based on a Bayesian learning algorithm, the {\sc BAT} package.  {\sc
  BAT} is based on the Markov Chain Monte Carlo technique, and
provides all the standard quantities such as best fit parameters,
goodness-of-fit, upper limits, etc.  In addition, {\sc BAT} provides a
sampling of the parameters of the model under study according to the
full posterior probability density function.  This allows for detailed
investigations of correlations between parameters, and the evaluation
of the probability density for any function of the parameters, without
approximations.  We believe this extra functionality will be very
useful in data analysis.

The {\sc BAT} package is coded in C++ and is available under \linebreak 
\url{http://mppmu.mpg.de/bat/}. A manual describing the use of {\sc BAT} is
also available at that location.

\section{Acknowledgements}
We would like to thank Giulio d'Agostini and Massimo Corradi for very
enlightening discussions. Special thanks also to Andrea Knue, an early
user of the {\sc BAT} package, and Daniel Greenwald.

\end{document}